\documentclass[preprint, aps, amsmath,amssymb,groupedaddress]{revtex4}
\usepackage{graphicx}
\pdfoutput=1
\usepackage{amsmath}
\usepackage{color}
\usepackage{float}

\begin{document}
\title{Direct evidence of strong local ferroelectric ordering in a thermoelectric semiconductor}

\author{Leena Aggarwal}
\affiliation{Department of Physical Sciences, Indian Institute of Science Education and   Research (IISER), Mohali, Punjab, India, PIN: 140306}
\author{Jagmeet S. Sekhon}
\affiliation{Department of Physical Sciences, Indian Institute of Science Education and   Research (IISER), Mohali, Punjab, India, PIN: 140306}
\author{Satya N. Guin}
\affiliation{New Chemistry Unit and International Centre for Materials Science, Jawaharlal Nehru Centre for Advanced Scientific Research (JNCASR), Jakkur, Bangalore, India, PIN: 560064}

\author{Ashima Arora}
\affiliation{Department of Physical Sciences, Indian Institute of Science Education and   Research (IISER), Mohali, Punjab, India, PIN: 140306}

\author{Devendra S. Negi}

\affiliation{New Chemistry Unit and International Centre for Materials Science, Jawaharlal Nehru Centre for Advanced Scientific Research (JNCASR), Jakkur, Bangalore, India, PIN: 560064}
\author{Ranjan Datta}

\affiliation{New Chemistry Unit and International Centre for Materials Science, Jawaharlal Nehru Centre for Advanced Scientific Research (JNCASR), Jakkur, Bangalore, India, PIN: 560064}



\author{{Kanishka Biswas}}
 
\email{kanishka@jncasr.ac.in}

\affiliation{New Chemistry Unit and International Centre for Materials Science, Jawaharlal Nehru Centre for Advanced Scientific Research (JNCASR), Jakkur, Bangalore, India, PIN: 560064}

\author{{Goutam Sheet}}
\email{goutam@iisermohali.ac.in}

\affiliation{Department of Physical Sciences, Indian Institute of Science Education and   Research (IISER), Mohali, Punjab, India, PIN: 140306}

\begin{abstract}

It is thought that the proposed new family of multi-functional materials namely the ferroelectric thermoelectrics may exhibit enhanced functionalities due to the coupling of the thermoelectric parameters with ferroelectric polarization in solids. Therefore, the ferroelectric thermoelectrics are expected to be of immense technological and fundamental significance. As a first step towards this direction, it is most important to identify the existing high performance thermoelectric materials exhibiting ferroelectricity. Herein, through the direct measurement of local polarization switching we show that the recently discovered thermoelectric semiconductor $AgSbSe_{2}$ has local ferroelectric ordering. Using piezo-response force microscopy, we demonstrate the existence of nanometer scale ferroelectric domains that can be switched by external electric field. These observations are intriguing as $AgSbSe_{2}$ crystalizes in cubic rock salt structure with centro-symmetric space group (Fm-3m) and therefore no ferroelectricity is expected. However, from high resolution transmission electron microscopy (HRTEM) measurement we found the evidence of local superstructure formation which, we believe, leads to local distortion of the centro-symmetric arrangement in $AgSbSe_{2}$ and gives rise to the observed ferroelectricity. Stereochemically active $5s^{2}$ lone pair of Sb can also give rise to local structural distortion, which creates ferroelectricity in $AgSbSe_{2}$.

\end{abstract}

\maketitle

Thermoelectric materials enable direct and reversible conversion of untapped heat into electrical energy and will play a significant role in the future energy management \cite{Bis,Bis1,Zhao,Pei,Hereman,Poudel}. Ferroelectric materials have a spontaneous electric polarization that can be reversed by the application of an external electric field. Ferroelectric materials have several applications such as in capacitors with tunable capacitance \cite{Tom,Stam}, non volatile memory devices \cite{Hu}, sensors \cite{Muralt}, transistors \cite{scott,Naber}, photovoltaic devices \cite{Yuan}, thermistor \cite{Lan} etc. Therefore, combining thermoelectric and ferroelectric properties in a single compound might lead to a new class of multiferroics namely the ferroelectric thermoeletrics that would undoubtedly be of immense fundamental and technological importance. In order to accomplish this, it is most important to first identify the existing high performance thermoelectric materials containing electrically active ferroelectric domains and investigate how ferroelectricity is coupled with thermoelectricity.

\begin{figure*}
\includegraphics[scale=0.5]{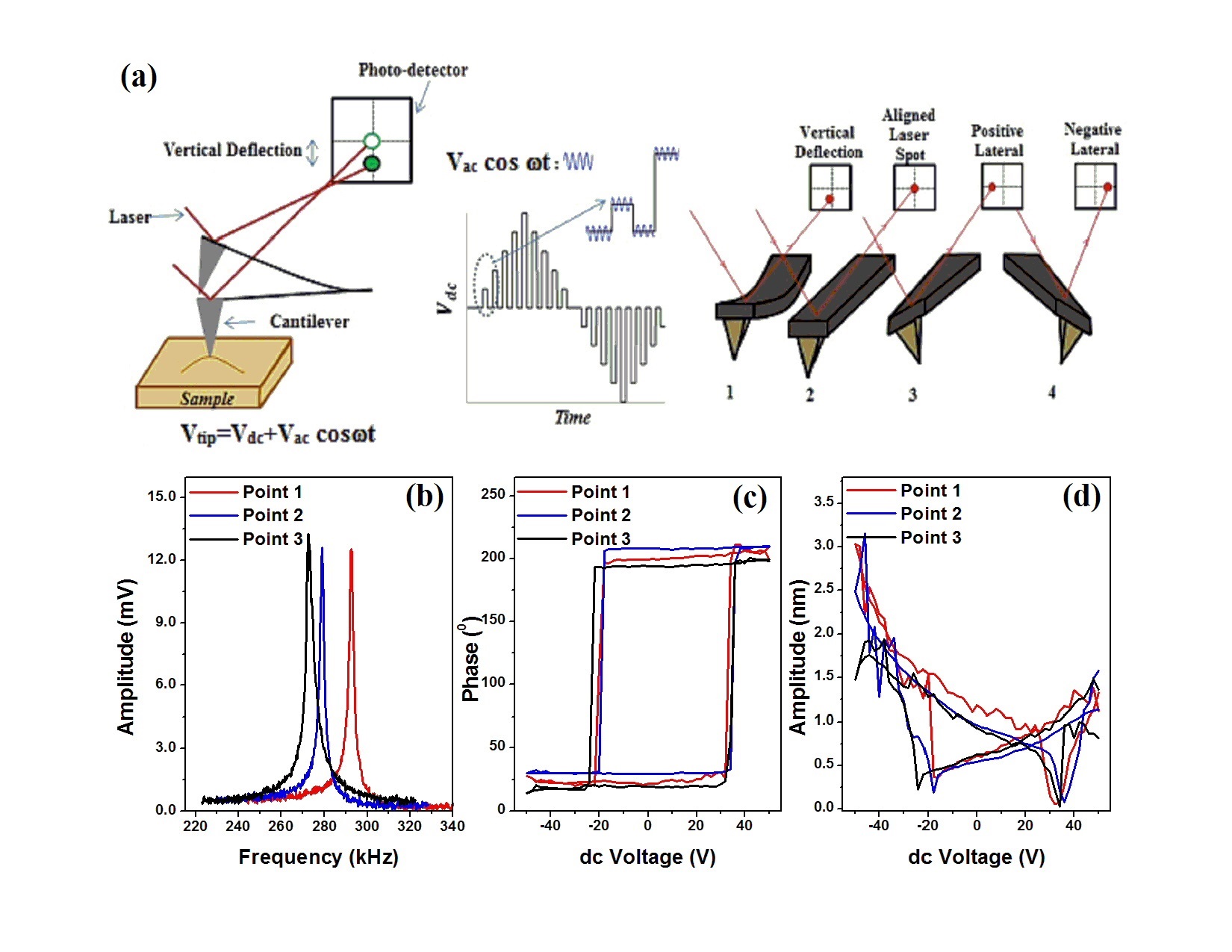}

\caption {Direct evidence of the ferroelectric and piezoelectric response of $AgSbSe_{2}$ using a conductive $Pt\backslash Ir$ coated tip studied by piezoresponce force microscopy (PFM). (a) Schematic of the PFM technique presenting the switching waveform in the DART PFM spectroscopic mode and the tip movement in Vector PFM mode. (b) Tuning of the conducting AFM cantilever in DART PFM mode prior to spectroscopic measurements. (c) PFM phase hysteresis loop and (d) butterfly loop measured at three different points in the ``off"-state.}
\end{figure*}

Although the direct evidence of ferroelectricity in thermoelectric materials through the imaging of ferroelectric domains is absent in the literature, few thermoelectrics have been shown to possess local electric polarization. For example, ferroelectric Aurivillius phase, $Bi_{4}Ti_3{O}_{12}$, exhibited enhanced thermoelectric behaviour \cite {Shen}. Single crystal of the relaxor ferroelectric oxide, $Sr_{1-x}Ba_{x} Nb_{2} O_{6}$, showed low thermal conductivity and high power factor \cite{Soonil Lee}. In the lead chalcogenides, which are known to be high performance thermoelectric materials, local ferroelectric distortion has been observed by pair distribution function analysis of temperature dependent neutron diffraction data \cite{Kanatzidis_local}. Using combination of inelelastic neutron scattering and first principle calculations, strong anharmonic coupling between ferroelectric transverse optic mode and longitudinal acoustic modes have been evidenced in $PbTe$, which has been shown to be important for low thermal conductivity \cite{Delaire}. In certain systems, it was found that the itinerant electrons responsible for electrical conductivity were strongly coupled with the ferroelectric polarization \cite{Soonil Lee,Kolo}. Therefore, in principle, by tuning ferroelectric properties it should also be possible to tune electrical properties, hence thermoelectric properties, in those systems \cite{Nina}. In such systems, the nano-meter scale ferroelectric domain boundaries could act as potential scatterers of the mid/long mean free path phonons thereby limiting the thermal conductivity. However, complete understanding of the correlation between ferroelectric and thermoelectric properties, so far, has remained elusive. This understanding is extremely important in order to be able to design novel materials with superior thermoelectric properties by hitherto unexplored techniques like engineering of ferroelectricity in semiconductors. Investigating the possibility of the existence of ferroelectric ordering in novel thermoelectric materials is, therefore, a major step forward to this direction.

Here, we present the existence of local ferroelectric ordering by piezoresponse force microscopy (PFM) in a new thermoelectric semiconductor $AgSbSe_{2}$. We observe signature of switching of local ferroelectric domains in response to a dc-voltage applied between a conducting atomic force microscope (AFM) cantilever and an oriented polycrystal of $AgSbSe_{2}$. From the measurement of local strain vs. voltage we also observe a clear butterfly loop which is a hallmark of piezoelectricity.

$AgSbSe_2$ crystallizes in cubic rock salt structure (space group, $Fm-3m$) with disorderd Ag and Sb positions. $AgSbSe_{2}$ is a p-type semiconductor that shows high thermoelectric figure of merit, $zT$ of $\sim 1.2$, when properly doped \cite {Guin, Guin1}. The high $zT$ was attributed to enhance electrical conductivity and low lattice thermal conductivity. The low lattice thermal conductivity arises due to the presence of high degree of anharmonicity in the $Sb-Se$ bond \cite {Morelli, Nielsen}. Our observation of the existence of strong local ferroelectricity is intriguing as $AgSbSe_{2}$ crystalizes in centro-symmetric space group $(Fm-3m)$ and therefore no ferroelectricity is expected. However, from high resolution transmission electron microscopy (HRTEM) measurement we found the evidence of local superstructure formation which, we believe, leads to local distortion of the centro-symmetric arrangement in $AgSbSe_{2}$ and gives rise to the observed ferroelectricity. Stereochemically active $5s^{2}$ lone pair of Sb can also give rise to local structural distortion, which creates ferroelectricity in $AgSbSe_{2}$.

\par
In order to investigate ferroelectric properties of  $AgSbSe_{2}$, we employed piezo-response force microscopy (PFM). A schematic diagram describing the experimental details has been provided in Figure 1(a). We engaged a conductive tip mounted on a conductive cantilever on the surface of a rectangular piece of $AgSbSe_{2}$.  For probing the piezoresponse of the microscopic region underneath the tip, an ac voltage $V_{ac}$ was applied between the tip and the sample mounted on a metal piece which was directly connected to the ground of the voltage source. The frequency of $V_{ac}$ was swept and the amplitude response of the cantilever in the contact mode was recorded in order to identify the contact resonance frequency of the cantilever on the sample. All the measurements presented in this paper were performed at the contact resonance frequency in order to achieve highest sensitivity. As shown in Figure 1(b), the contact resonance frequency varied between $270$ kHz to $300$ kHz during our measurements. 

For PFM spectroscopy, the ac voltage $V = V_{ac} cos\omega t$ is mixed with a dc voltage $(V_{dc})$ and the sum $ V_{tip} = V_{dc} + V_{ac} cos\omega t$ is applied between the tip and the sample. If the sample is piezoelectric, the area underneath the tip will be deformed due to the application of the electric field. Since the applied field has a periodic component, the amplitude of deformation can be written as $ A = A_0 + A_\omega cos(\omega t + \phi)$, where $\phi$ is the phase difference between the applied field and the amplitude response. $\phi$ gives information about the electric polarization direction below the tip. Depending on the direction of the applied $V_{dc}$, the polarization underneath the tip switches direction. This is expected to give rise to a $180^0$ shift in $\phi$ vs $V_{dc}$ curve as passes through $V_{dc} = 0$. In general, for ferroelectric switching, the phase $\phi$ switches hysteretically and shows a finite coercive voltage. In Figure 1(c) we show the hysteresis loops obtained at three different points on $AgSbSe_2$. The hysteretic $180^0$ phase switching is clearly visible. For piezoelectric samples, when amplitude $A_\omega$ is plotted against the sweeping $V_{dc}$, the curve is expected to be hysteretic resembling a butterfly. This is called a ``butterfly loop" which is traditionally considered to be a hallmark of piezoelectricity. The representative butterfly loops measured at three different points on $AgSbSe_2$ are shown in Figure 1(d) (also see the supplementary material).

Observation of hysteretic phase switching and ``butterfly loops" indicate that $AgSbSe_2$ has ferroelectric and piezoelectric properties. However, it should also be noted that such hysteresis in phase and amplitude may also arise from electrostatic and electrochemical effects \cite{Jagmeet}. In order to minimize the role of the electrostatic effects, all the measurements were performed following SS-PFM (switching spectroscopy piezoresponse force microscopy) pioneered by Jesse $et.al.$\cite {Jesse, Jesse1}. In this method, instead of sweeping $V_{dc}$ continuously, it is applied in sequence of pulses and the phase and amplitude measurements are done in the ``off"-states of the pulses. In the supplementary material \cite{EPAPS} we have shown the difference between the ``on"-state and the ``off"-state measurement results. The difference is significant indicating that the electrostatic effects have been minimized in the ``off"-state measurements. The possibility of electrochemical reactions under the tip is ruled out by topographic imaging after the spectroscopic measurements, where we do not observe any topographic modification that is usually expected to result from tip-induced electrochemical processes \cite{Jagmeet}. However, ruling out the above-mentioned possibilities with absolute certainty may not be possible by spectroscopic measurements alone. In addition to the hysteretic switching effects, the observation of domains on the sample surface would be unambiguous proof of ferroelectricity. We present the observation of ferroelectric domains by PFM imaging in the following section.

\begin{figure}
\begin{center}
\includegraphics[width=\textwidth]{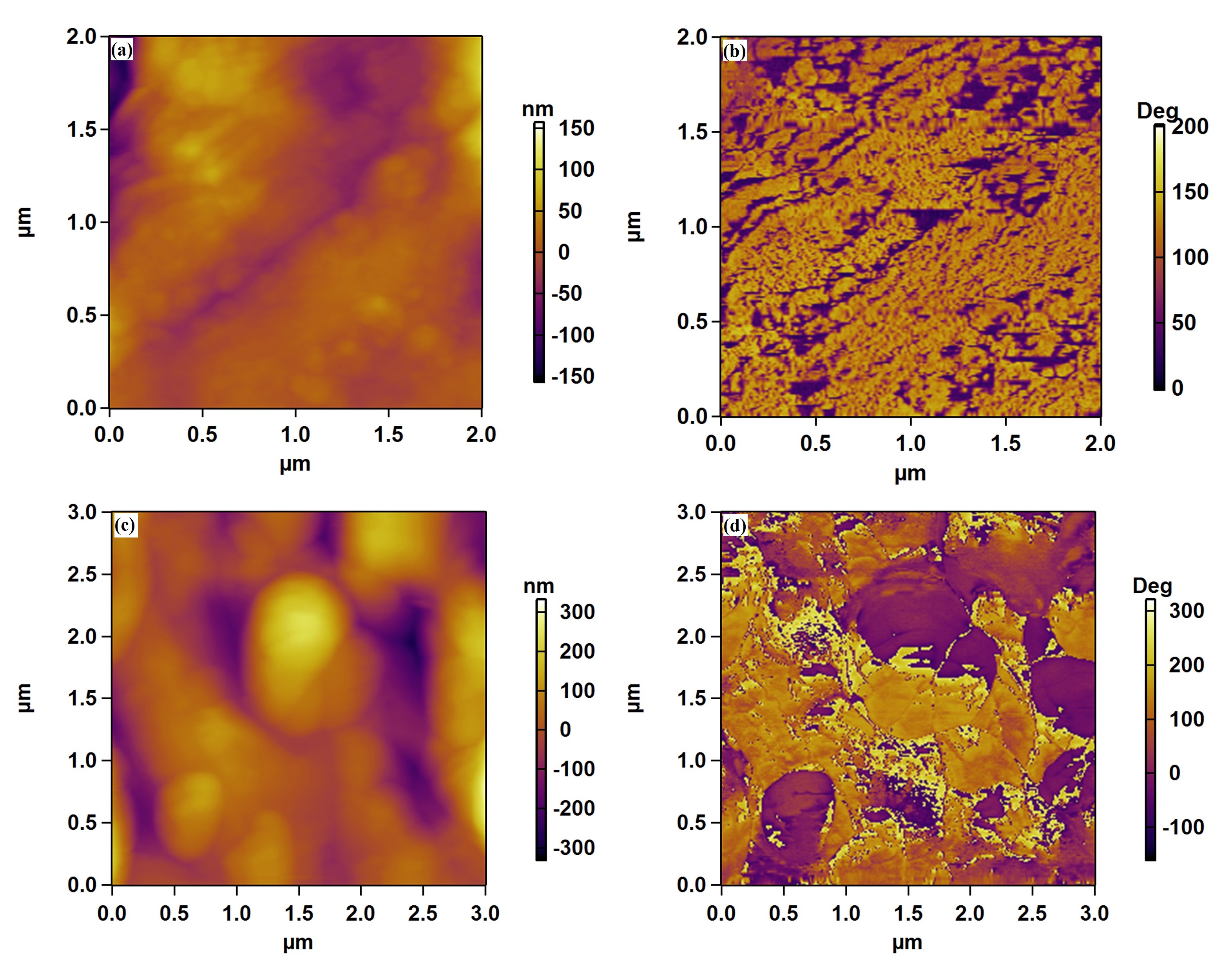}
\end{center}

\caption{(a) Topography of a $2 \mu m \times 2\mu m$ area on the sample. (b) PFM phase image showing nanometer scale domains in the same area as in (a). (c) Topography of a $3\mu m\times3 \mu m$ area on the sample. (d) Lateral PFM phase image showing nanometer scale domains in the same area as in (c).} 
\end{figure}

PFM imaging was done by bringing the tip in contact with the sample, applying the voltage $V_{ac}$ on the tip and scanning the tip on the sample while keeping the deflection of the cantilever constant. If the electric polarization is opposite in two different domains, the measured value of $\phi$ varies by $180^0$ between the two domains. By plotting the magnitude of $\phi$ as a function of the position of the tip, a PFM image exhibiting the distribution of ferroelectric domains on the sample surface is constructed. During PFM imaging the resonance frequency might shift as the cantilever rubs against the surface and interacts differently with the sample at different points. In order to track the contact-resonance in real time the measurements were done in so-called DART (dual ac resonance tracking) mode \cite{Rod}. In Figure 2(a) we show the topographic image of a $2\mu m$ x $2 \mu m$ area. In Figure 2(b) we show the phase image corresponding to the same area. The phase image shows regions with high (bright) and low (dark) values of $\phi$. From the bright to the dark region the phase $\phi$ shifts by $180^o$ indicating that the bright and dark regions are respectively `up' polarized and `down' polarized ferroelectric domains.

In the imaging method described above only the vertical amplitude and phase response on the photodiode are recorded. The image thus obtained gives information about the components of electric polarization along the directions parallel and antiparallel to the axis of the tip. By probing the lateral deflection of the cantilever it is also possible to image the domains where the polarization axis remains on the plane of the sample surface. The images of the lateral domains of a  $3 \mu m$ x $3 \mu m$ area (topography shown in Figure 2(c)) is presented in Figure 2(d). Bright and dark regions comprising of domains of different electric polarization are clearly visible in the image. There is absolutely no correlation between the topographic and the phase image confirming that the phase image is not affected by the topography.

From the above results it can be inferred that $AgSbSe_2$ has local ferroelectric properties. Since  $AgSbSe_2$ crystallizes in the centro-symmetric structure, the observation of ferroelectricity is surprising. It is therefore necessary to understand the origin of ferroelectricity in $AgSbSe_{2}$. 

$AgSbSe_2$ is a member of cubic $I-V-VI_{2}$, where I = Cu, Ag, Au or alkali metal; V = As, Sb, Bi; and VI = Se, Te. In  $AgSbSe_2$, the valence electronic configuration of Sb is $5s^{2}5p^{3}$, where only $5p^{3}$ electrons take part in the formation of bonds with the Se valence electrons, while the beguiling $5s^{2}$ electrons of Sb form lone pairs. Such $ns^{2}$ lone pairs electrons are often found to effective to distort the local structure, resulting in local dipole in the global centro-symmetric structure \cite{Kanatzidis_local,Ses, Wag, Neat, Rao}. Thus, lone pair induced local distortion in the structure may gives rise to ferroelectric properties in $AgSbSe_2$.

\
\begin{figure}
\includegraphics[width=\textwidth]{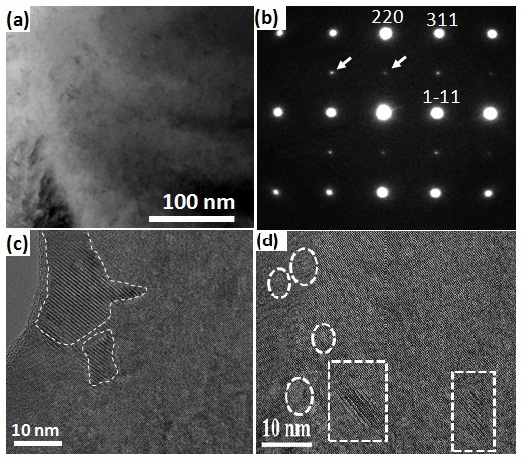}
\begin{center}
\caption{Nanoscle architecture of $AgSbSe_{2}$. (a) Low magnification TEM image of $AgSbSe_2$. (b) Fast Fourier transform (FFT) pattern of (a). Arrows in (b) show the weak superstructure spots. (c, d) High resolution TEM images of $AgSbSe_2$, dotted portion showing nanoprecipitates with the doubling of lattice parameter compared to matrix.} 

\end{center}
\end{figure}

In order to investigate the nanoscale architectures, we have performed transmission electron microscopic (TEM) investigation on pristine $AgSbSe_2$ (Figure 3(a)). Although earlier powder X-ray diffraction measurements of $AgSbSe_{2}$ indicated that Ag and Sb position were disordered in the cation site  of the NaCl-type structure \cite{Guin}, careful TEM studies revealed the evidence of local $Ag/Sb$ ordering in nanoscale regions distributed throughout the $AgSbSe_2$ sample. We have also observed weak superstructure spots in the electron diffraction along the $<110>$ and $<201>$ direction (arrows in Figure 3(b)), which resulted due to ordering of Ag and Sb atoms in $AgSbSe_{2}$. Figure 3(c) and Figure 3(d) depict representative high resolution TEM (HRTEM) images of $AgSbSe_{2}$ at two different locations of the sample. Close inspection of the images reveal the presence of two types of region in the sample. The dotted marked region in the image are nanodomains with doubled lattice parameter (Figure 3(c)), which might have formed due to local cation ordering in the $AgSbSe_{2}$. Similar ordering of cations in $AgSbTe_{2}$ had been earlier predicted by first-principle calculations for ground state structure \cite{Hoa, Bar} and were recently observed by inelastic neutron scattering and TEM investigations in $AgSbTe_{2}$ \cite{Ma}. Recent electrical transport measurements on cubic $AgBiS_{2}$ nanocrystals also indicated the presence of cation ordering near room temperature \cite{Guin2}. The local cation ordering in the nanoscale regime might alter the structure making it non-centrosymmetric locally, which in turn could be responsible for the observed local ferroelectricity in $AgSbSe_{2}$.

It was suggested in the past that ferroelectric domain-walls might scatter phonons causing a reduction of $k_{lat}$ \cite{Wei}. The effect of ferroelectric domains on the thermoelectric properties may be further investigated by artificially designing the domain structures by nanostructuring and/or by fabricating epitaxial thin films on different substrates \cite{Jang}.

In conclusion, we have performed piezoresponce force microscopy on $AgSbSe_2$ and observed the signature of strong local ferroelectricity and piezoelectricity. From nanoscale structural analysis we show that natural formation of superstructure nanodomains in the sample makes the local crystal structure to be non-centrosymmetric, thereby inducing ferroelectricity. The local dipole could also be created due to the local structural distortion caused by $5s^{2}$ lone pair of Sb. Nano-meter scale ferroelectric domains were clearly imaged by PFM. The naturally occurring ferroelectric domain-walls might act as effective scatterers for phonons and reduce the thermal conductivity of the materials. Based on the observation reported in this Letter, enhancing thermoelectricity by ferroelectric domain engineering might emerge as a promising field of research. Further, it would be interesting to search for ferroelectricity in other known $I-V-VI_{2}$ thermoelectric semiconductors.\\

GS and KB acknowledge the research grant of Ramanujan Fellowship from Department of Science and Technology (DST), India for partial financial support. KB also acknowledges Sheik Saqr laboratory for partial funding.

\end{document}